\documentclass[aps,pra,twocolumn,groupedaddress]{revtex4}
\usepackage{mathrsfs}
\usepackage{amsmath}
\usepackage{amssymb}
\usepackage{revsymb}
\usepackage{graphicx}
\usepackage{mathrsfs}
\usepackage{psfrag} 
\usepackage[utf8]{inputenc}

\usepackage{color} 
\usepackage{ulem}  
\newcommand{\bra}[1]{\langle\,{#1}\, |}
\newcommand{\ket}[1]{|\,{#1}\,\rangle}
\newcommand{\braket}[2]{\mbox{$\langle\,{#1}\, | \,{#2}\,\rangle$}}

\newcommand{\Real}{\mbox{Re}}

\newcommand{\ElTransE}{ \varepsilon}

\newcommand{\LichtFr} {\nu}  
    
\newcommand{\V}{V}

\newcommand{ \VibFr} {\omega}

\newcommand{ \vc} { \kappa}

\newcommand{\HamAggEl}{H_{\rm sys}}

\newcommand{\HamAggBath}{H_{\rm env}}
\newcommand{\HamAggInt}{H_{\rm int}}

\newcommand{\AnzMon}{N}

\newcommand{\aDestroy}{a}
\newcommand{\aPlus}{a^{\dagger}}

\newcommand{\BathCor}{\alpha}

\newlength{\mylenunit}
\begin{document}


\title{Anomalous strong exchange narrowing in excitonic systems}

\author{Jan Roden}
\email{roden@mpipks-dresden.mpg.de}

\author{Alexander Eisfeld}
\affiliation{Max Planck Institute for the Physics of Complex Systems, N\"othnitzer Str.~38, D-01187 Dresden, Germany}

\date{\today}

\begin{abstract}
We investigate theoretically  the phenomenon of exchange narrowing in the absorption spectrum of a chain of monomers, which are coupled via resonant dipole-dipole interaction.
The individual (uncoupled) monomers exhibit a broad absorption line shape due to the coupling to an environment consisting of a continuum of vibrational modes. 
Upon increasing the interaction between the monomers, the absorption spectrum of the chain narrows. 
For a non-Markovian environment with a Lorentzian spectral density, we find a
narrowing of the peak width (full width at half maximum (FWHM)) by a factor $1/N$, where $N$ is the number of monomers. 
This is much stronger than the usual $1/\sqrt{N}$ narrowing.
Furthermore it turns out that for a Markovian environment no exchange narrowing at all occurs.
The relation of different measures of the width (FWHM, standard deviation) is discussed.
\end{abstract}


\maketitle

\section{Introduction}
Due to interactions with an environment the absorption lines of electronic transitions can be considerably broadened.
This broadening might be due to the fact that the individual (two-level) systems see different environments (static disorder) or that the excitation couples to time-dependent fluctuations or vibrational modes of the environment (dynamic disorder). 
In this work we refer to the individual two-level systems as monomers.

When there is exchange of excitation between the  monomers (e.g.\ due to transition dipole-dipole interaction), excitonic states are formed.
In the case of vanishing coupling to the environment, this leads to the
formation of a band of eigenstates having a
width proportional to the interaction strength between the monomers.
Due to selection rules not all of the states absorb.
For a parallel arrangement of the transition dipoles of the monomers, the
dominant absorption will be into states at the band edge. 

One might expect that for the coupled monomers the interaction with the environment will lead to a similar broadening  as in the case of uncoupled monomers. 
However, for strong coupling between the monomers the aggregate absorption line is considerably narrowed compared to that of the uncoupled monomers.
This exchange narrowing effect occurs for example in paramagnetic resonance~\cite{AnWe53_269_}, in the excitonic line
shapes of disordered solids~\cite{KoeJaRe89_451_}, or for the J-band of molecular aggregates~\cite{Kn84_73_,FiKnWi91_7880_,Ma93_225_,WaEiBr08_044505_}.
To gain a better understanding of the narrowing mechanism, many numerical and analytical studies have been performed~\cite{WaEiBr08_044505_,Kn84_73_,FiKnWi91_7880_,Ma93_225_,BlSi78_3589_,ScTo82_1528_,WuKn98_63_}.
From these studies it has emerged that if there are no correlations between the individual environments of the different monomers, in the case of a one-dimensional arrangement of $N$ monomers the line shape narrowing is usually of the order of $1/\sqrt{N}$ compared to the width of the line shape of the non-interacting monomers and occurs if the  coupling between the monomers is much larger than the coupling to the environment.
However, recently it has emerged~\cite{EiBr06_113003_,WaEiBr08_044505_,VlMaKn09_205121_,Ei_Vlaming_PRL} that for certain disorder distributions a totally different behaviour can be found. 
For example in the case of a Lorentzian distribution (i.e.\ a Lorentzian absorption line shape) there is no narrowing at all~\cite{EiBr06_113003_,WaEiBr08_044505_,VlMaKn09_205121_} and for distributions, which fall off slower than a Lorentzian, even a broadening can occur~\cite{Ei_Vlaming_PRL}.

In the present paper we will not consider static disorder, but focus on the coupling of the excitation to vibrational modes of the environment.
We use a commonly adopted model in which the excitation of a monomer is linearly coupled to harmonic modes of its environment~\cite{ScFi84_269_,WaEiBr08_044505_,MaKue00__,SeWiRe09_13475_,Sp09_4267_,GuZuCh09_154302_}.
This model, restricted to only a single environmental mode, is often used to describe the influence of a high energy vibrational mode in aggregates of organic dye molecules~\cite{EiBrSt05_134103_,KoHaKa81_498_,SeWiRe09_13475_}. 
For one mode the resulting absorption spectrum of non-interacting monomers
consists of delta-functions spaced by the frequency
$\Omega$ of the mode and the weight of the delta-functions, i.e.\ their ``height'', follow a
Poisson distribution~\cite{WaEiBr08_044505_,Sp09_4267_,MaKue00__} whose standard deviation (width) is determined by the coupling strength between the
electronic excitation and the vibrational mode (note that a Poisson distribution is completely determined by the standard deviation).
For a chain of $N$ monomers one then finds that in the limit of very strong
interaction (i.e.\ the interaction between the monomers is much larger than
the interaction between electronic excitation and vibrations) the absorption
spectrum is again a Poissonian consisting of delta-functions spaced by $\Omega$, but with a standard deviation decreased by a factor $1/\sqrt{N}$.
This leads to an increase of the intensity of the delta-function at lowest energy w.r.t.\ the delta-functions at higher energies.
For more than one mode a similar pattern is found: the monomer line shape
consists of a convolution of the Poissonians of the different modes,
  and consequently its variance is given by the sum of the variances of the
  individual Poisson distributions. 
For the chain the standard deviation of each Poissonian
is diminished by a factor $1/\sqrt{N}$
so that also the standard deviation of the whole peak structure is
  diminished by a factor $1/\sqrt{N}$. 
When the density of modes becomes large (continuous), the resulting absorption spectrum becomes continuous, too. Often this spectrum is then dominated by a single band/peak which possesses a more or less pronounced shoulder.
In such a case it is common to characterize the the width of this peak by its full width at half maximum (FWHM). In the following we will also use this measure.
In the conclusion we will discuss the relation between different measures for the width of an absorption spectrum w.r.t.\ exchange narrowing.  

In this work we investigate the transition from the monomer line shape to the
exchange narrowed line shape of the one-dimensional aggregate, considering a
continuous spectral density of the environment, which leads to a
  continuous absorption lineshape of the monomer/aggregate.
We will focus in particular on the FWHM of the dominant absorption peak. 
We will find, that in analogy to the case of Lorentzian static disorder also in the case of a Lorentzian monomer absorption line shape (corresponding to a Markovian environment) there is no exchange narrowing.
On the contrary, in the case of a Lorentzian spectral density (which leads to an asymmetric structured monomer line shape) an enhanced narrowing is found.

The paper is organized as follows:
In Section~\ref{sec:model}, the Hamiltonian we will consider is introduced. 
In Section~\ref{sec:abs_spec}, the method to calculate the zero temperature absorption spectrum is briefly reviewed. 
The exchange narrowing is investigated in Section~\ref{sec:results}.
We conclude in Section~\ref{sec:conclusions} with a discussion of our findings.
In the appendix, an analytical formula for the monomer absorption spectrum is given.

\section{The model Hamiltonian}
\label{sec:model}
We consider an assembly of $N$  two-level systems (i.e.\ one two-level system per monomer). 
The transition energy between the ground state $\ket{\phi_n^g}$ and the excited state $\ket{\phi_n^e}$ of monomer $n$ is denoted by $\ElTransE_n$.
We will consider absorption from the state where all monomers are in their ground state.
The state in which monomer $n$ is  excited and all other 
monomers are in their ground state is denoted by $\ket{\pi_n}$. 
In the following we will not consider states in which the whole system has more than one excitation.
 Expanding the Hamiltonian of the interacting monomers with respect to the ``one-exciton'' states $\ket{\pi_n}$  gives  
\begin{equation}
\label{eq:Ham_sys}
\HamAggEl=\sum_{n,m=1}^N\Big(\ElTransE_n\delta_{nm}+ 
\V_{nm}\Big)\ket{\pi_n}\bra{\pi_m},
\end{equation}
where the interaction $V_{nm}$ describes exchange of excitation between monomer $n$ and $m$.
Taking also the interaction with the environment into account the total Hamiltonian in the one-exciton subspace can be written as~\cite{MaKue00__}
\begin{equation}
\label{eq:ham_tot}
H=\HamAggEl+\HamAggInt+\HamAggBath.
\end{equation}
The environment of each monomer is taken to be a set of harmonic modes
described by the Hamiltonian 
\begin{equation}
\HamAggBath=\sum_{n=1}^{\AnzMon}\sum_{\lambda} \VibFr_{n\lambda} \aPlus_{n\lambda}  
\aDestroy_{n\lambda}.
\end{equation}
Here $\aDestroy_{n\lambda}$ denotes the annihilation operator of 
mode $\lambda$ of monomer $n$ with frequency $\VibFr_{n\lambda}$.
Note that each monomer possesses its own environment which is
  independent of the environments of the other monomers.
The coupling of an excitation on a monomer to the
  modes of its local environment is assumed to be linear and the corresponding interaction Hamiltonian is given by
\begin{equation}
\HamAggInt=-\sum_{n=1}^{\AnzMon}\ket{\pi_n}\bra{\pi_n} \sum_{\lambda} \vc_{n\lambda} (\aPlus_{n\lambda}  
+\aDestroy_{n\lambda}),
\end{equation}
where the coupling constant $\vc_{n\lambda}$ specifies the
  strength of the coupling of the excitation on monomer $n$ to the mode $\lambda$ with frequency $\omega_{n\lambda}$ of the local environment.
This interaction is conveniently described by introducing the spectral density~\cite{MaKue00__}
\begin{equation}
\label{spec_dens}
J_n(\omega)=\sum_{\lambda} |\vc_{n\lambda}|^2 \delta (\omega-\omega_{n\lambda})
\end{equation}
of monomer $n$.
In this work we will consider $J_n(\omega)$ to be a continuous function.
The spectral density is closely related to the correlation function of the environment (bath correlation function).
We restrict the discussion in this work to the case of zero temperature where the bath correlation function $\BathCor_{n}(\tau)$ reduces to
\begin{equation}
\label{eq:BathCor}
\BathCor_{n}(\tau)= \int d\omega\ J_n(\omega)\, e^{-i \omega \tau}.
\end{equation}
For a continuous spectral density $J_n(\omega)$ the bath correlation function is decaying to zero on a time scale determined by the width of $J_n(\omega)$.
When the environment has no memory, i.e.\ $\BathCor_{n}(\tau) \propto \delta(\tau)$, it is termed Markovian, otherwise it is non-Markovian.

\section{The absorption spectrum}
\label{sec:abs_spec}

At zero temperature, initially the aggregate is in a state $\ket{g}$ in which all the monomers are in their ground state $\ket{\phi_n^g}$ and also all environmental modes are in their ground state, i.e.\
\begin{equation}
  \ket{g}= \ket{g_{\rm vib}}\prod_{n=1}^N\ket{\phi_n^g},
\end{equation}
where $\ket{g_{\rm vib}}$ denotes the product of the vibrational ground states of all modes.
For simplicity we take all transition dipoles of the monomers to be identical.
Then the absorption strength for light with frequency 
 $\LichtFr$  is given by~\cite{MaKue00__,La52_1752_}
 \begin{equation}
\label{eq:abs_spek_mit_tkorr_fkt}
   A(\LichtFr)= \Real \int_0^{\infty}dt\ {\rm e}^{i\LichtFr t}\bra{\Psi_0}{\rm e}^{-iHt}\ket{\Psi_0}
 \end{equation}
with 
\begin{equation}
 \ket{\Psi_0}=\ket{\psi_0}\ket{g_{\rm vib}}
\end{equation}
and 
\begin{equation}
\label{def_el_psi_0}
  \ket{\psi_0}=\frac{1}{\sqrt{N}}\sum_{n=1}^N\ket{\pi_n}.
\end{equation}
The calculation of the aggregate absorption spectrum poses serious problems in the general case, due to the complex structure of the coupling to the environmental oscillators.
In certain cases however, it is possible to solve the problem numerically or even analytically.
In the following we will first consider the Markovian case, where we show analytically that no narrowing occurs.
   To demonstrate the effect of anomalous strong narrowing we will concentrate on a spectral density which is a single Lorentzian. For such a spectral density we are able to calculate the absorption spectrum for quite large aggregates numerically exact, using  the method described in Ref.~\cite{ZOFE_PM}. 

\section{Investigation of the exchange narrowing}
\label{sec:results}

To investigate the appearance of exchange narrowing we choose a one
dimensional arrangement of the monomers and take the  transition energies to
be equal, $\ElTransE_{n}=\ElTransE$.  
We introduce periodic boundary conditions (except for the dimer)
 and take only coupling between nearest neighbors into account, which we assume to be equal for all neighboring monomers, i.e.\ $V_{n,n+1}=V$ for all $n$. 
For simplicity (as already mentioned in the previous section) we also take all the transition dipole moments to be identical.
For this arrangement, the absorption spectrum, in the limit of vanishing coupling to the environment, simply consists of one absorption line located at the band edge of the exciton-band~\cite{Kn84_73_,EiBr06_376_,Ma93_225_,FiKnWi91_7880_}
\begin{equation}
\label{eq:Abs_rein_elektronisch}
A(\LichtFr)=  \delta\big(\LichtFr- (\ElTransE+C)\big),
\end{equation}
where $C=V$ in the case of a dimer and $C=2V$ for longer aggregates.

\subsection{Markovian case}

For a Markovian environment, one has $\BathCor_{n}(\tau)\propto \delta(\tau)$, specifically we use $\BathCor_{n}(\tau)=2 \Gamma \delta(\tau)$.
In this case, one can show~\cite{YuDiGi99_91_,RoEiWo09_058301_,ZOFE_PM} that $\bra{\Psi_0}{\rm e}^{-iHt}\ket{\Psi_0}$ needed in Eq.~(\ref{eq:abs_spek_mit_tkorr_fkt}) is given by $\braket{\psi_0}{\psi(t)}$ with $\ket{\psi_0}$ from Eq.~(\ref{def_el_psi_0}) and where the purely electronic state $\ket{\psi(t)}$ can be obtained from solving
\begin{equation}
 \label{Schr_Markov}
 \partial_t \ket{\psi(t)}=\left(-i\HamAggEl - \Gamma\right)\ket{\psi(t)}
\end{equation}
with the initial condition $\ket{\psi(t)}=\ket{\psi_0}$.
One can easily solve Eq.~(\ref{Schr_Markov}) and use Eq.~(\ref{eq:abs_spek_mit_tkorr_fkt}) to obtain the absorption line shape
\begin{equation}
  \label{eq:Markov_lineshape}
 A(\LichtFr)= \frac{1}{2\pi}\,\frac{\Gamma}{\big(\LichtFr- (\ElTransE+C)\big)^2 +\Gamma^2}.
\end{equation}
This is a Lorentzian line shape centered at $(\ElTransE+V)$ in the case of a dimer and at $(\ElTransE+2V)$ for $N\ge 3$. 
The  width $\Gamma$ is independent of the interaction $V$ between the monomers
and  of the number $N$ of the monomers, i.e.\ no narrowing occurs. 
A similar result has previously also been observed for the case of Lorentzian static disorder~\cite{EiBr06_113003_,WaEiBr08_044505_,VlMaKn09_205121_}.

\subsection{Lorentzian spectral density}
\label{sec_lorentz_spec_dens}

For the environment we consider a continuum of frequencies  so that the spectral density $J_n(\omega)$ becomes a smooth function. 
We will consider a Lorentzian spectral density
\begin{equation}
\label{lorentz_spec_dens}
 J_n(\omega)= \frac{1}{\pi}\Omega^2 X\frac{\gamma}{(\omega-\Omega)^2+\gamma^2}.
\end{equation}
Note that in the limit $\gamma \rightarrow 0$ this spectral density becomes a delta function peaked at the frequency $\Omega$. 
This leads to the well investigated model, in which each monomer possesses one undamped vibrational mode of frequency $\Omega$ that couples to the electronic excitation~\cite{ScFi84_269_,WaEiBr08_044505_,SeWiRe09_13475_,Sp09_4267_,KoHaKa81_498_,FuGo61_1059_}.
The quantity $X$ appearing in Eq.~(\ref{lorentz_spec_dens}) is the Huang-Rhys factor~\cite{MeOs95__} that indicates the strength of the coupling between electronic excitation and the vibrational mode.
The width $\gamma$ of the Lorentzian spectral density is the magnitude of the damping of the vibrational motion.

\subsubsection{The monomer spectrum}

 \begin{figure*}[btp]
\includegraphics[width=0.48\mylenunit]{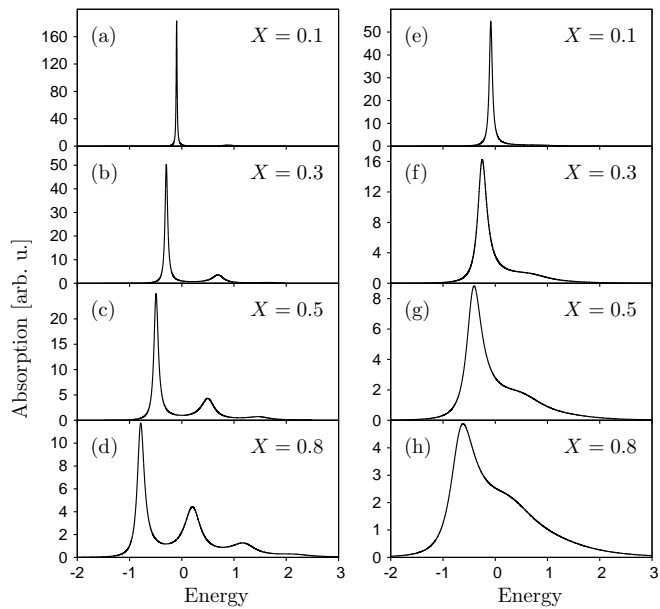}
 \caption{\label{fig:monomer} Absorption spectrum of a monomer (i.e.\ $V=0$) for various values of $X$ (indicated on the figures) and for $\gamma=0.1$ (left column) and $\gamma=0.4$ (right column).
The frequency $\Omega$, at which the Lorentzian spectral density is centered, is taken as the unit of energy (throughout the paper).}
 \end{figure*}

For a Lorentzian spectral density the monomer absorption line shape (i.e.\ for $V=0$) can be calculated analytically, as is done in the appendix.
In Figure~\ref{fig:monomer} the resulting monomer spectrum is shown for various parameters $X$ and $\gamma$. 
Here and throughout the paper the frequency $\Omega$, at which the Lorentzian spectral density is centered, is taken as the unit of energy.
As the zero of energy we choose the monomer transition energy $\ElTransE$, where the mean of the monomer spectrum is located.
In the left column of Fig.~\ref{fig:monomer} all spectra are for a width $\gamma=0.1$ and $X$ increases from top to bottom. 
For increasing $X$ the coupling of the electronic excitation to the environment becomes stronger resulting in the appearance of a vibrational progression. 
For a larger $\gamma$, i.e.\ a larger width of the spectral density, all peaks
of the monomer spectrum become broader, as can be seen in the right column of
Fig.~\ref{fig:monomer} where $\gamma=0.4$. 
Note, that for small Huang-Rhys factors ($X\lesssim 0.3$) the monomer spectrum is dominated by one peak.

\subsubsection{The narrowing of the aggregate spectrum}

\begin{figure*}
\includegraphics[width=0.95\mylenunit]{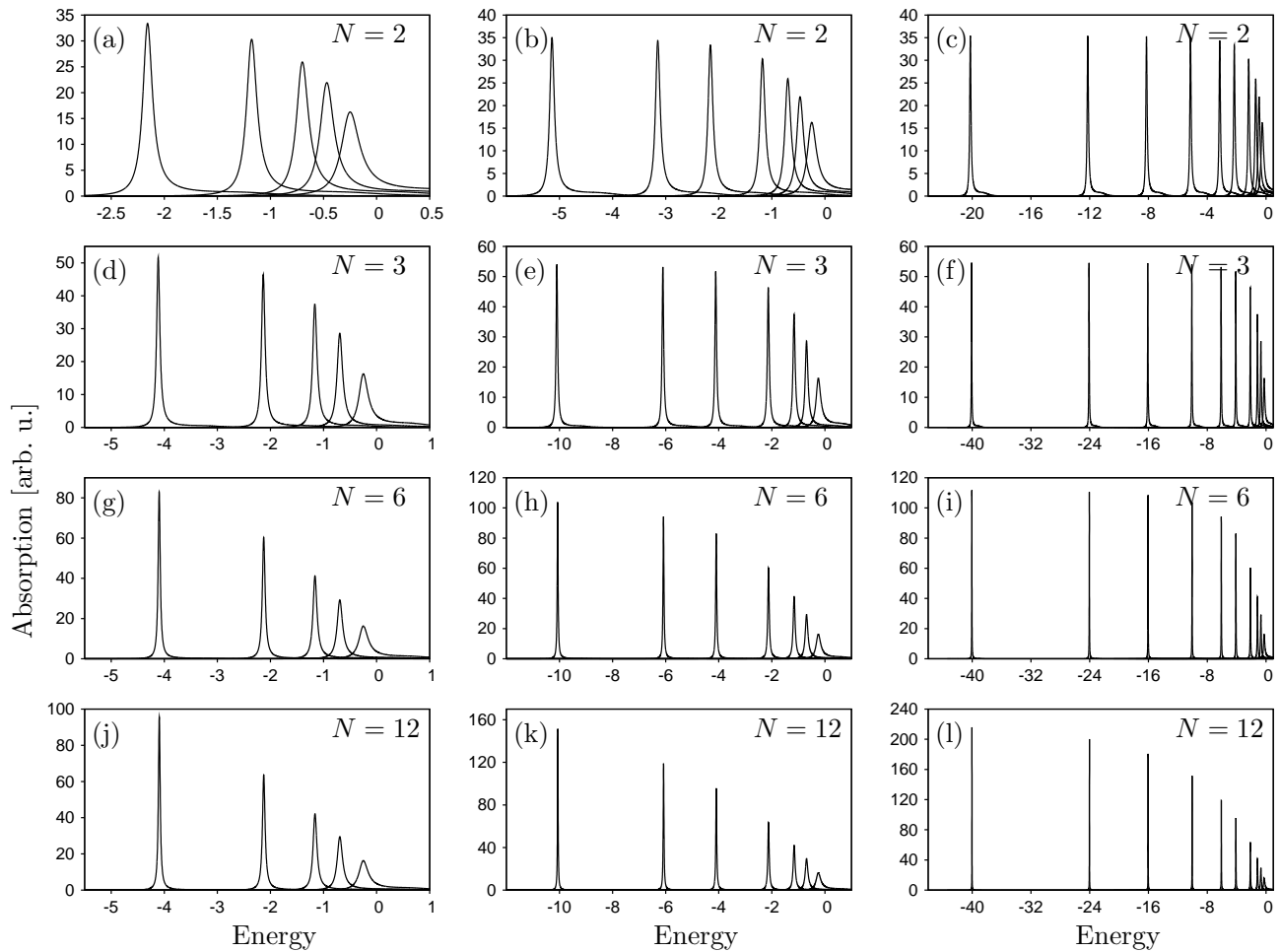}
\caption{\label{fig:gesamt}
Absorption spectra of different $N$-mers ($N$ indicated on the figures) for various $V$; 
the corresponding monomer spectrum is that of Fig.~\ref{fig:monomer}f, which is shown here again for comparison (right most spectrum in each plot).
In the first column the values of $V$ are (from right to left): $V= 0, -0.25, -0.5, -1, -2$.
In the second column in addition to the values of the first column also $V= -3, -5$ are shown.
Third column: In addition spectra for $V= -8, -12, -20$ are shown.
Note the different ranges of the abscissa of the individual plots. }
\end{figure*}

As an example, in the following investigations we will consider the case of the monomer spectrum Fig.~\ref{fig:monomer}f, obtained for $X=0.3$ and $\gamma=0.4$.
We now explore how this line shape narrows upon increasing the coupling strength $|V|$ between the monomers and examine its dependence on the number $N$ of monomers.
In Figure~\ref{fig:gesamt} aggregate spectra are shown for different $N$ and for various values of the coupling $V$.
For comparison, the corresponding monomer spectrum of Fig.~\ref{fig:monomer}f is also shown (right most curve in each plot).
For all spectra we chose $V<0$ (in molecular aggregates such a negative coupling leads to the formation of the famous J-band \cite{EiBr06_376_}). 
The larger the value of $|V|$, the more the spectrum is shifted to lower energies.
This is in accordance with a sum rule that states that the mean of the absorption line shape is shifted by $C=2V$ (and $C=V$ for the dimer, i.e.\ $N=2$, respectively) with respect to the mean of the monomer line shape~\cite{BrHe70_1663_,Ei07_321_}.
From another sum rule it follows that the area of the absorption line shape is independent of $V$. 
Thus the increase of the height of the absorption peak with increasing $|V|$ already indicates its narrowing.
Going from the first to the third column of Fig.~\ref{fig:gesamt} the maximal
$|V|$ that is shown is increased, leading to a shift of the peak position to
lower energies.

To investigate the narrowing with increasing $|V|$ in detail, consider first Fig.~\ref{fig:gesamt}a, where dimer spectra ($N=2$) are shown for relatively small dipole-dipole interaction ($|V|\le 2$). 
One sees, that upon increasing $|V|$  the width of the absorption spectrum becomes gradually narrower. 
This narrowing takes place quite rapidly for $|V|<1$ and  then slows down. 
For $V\approx -2$ the spectrum has nearly obtained its asymptotic form and does not change appreciably upon increasing $|V|$ further. 
This can be seen clearly in Figs.~\ref{fig:gesamt}b and~c, where dimer spectra for larger $|V|$ are shown.

In addition to the dimer case also spectra for $N=3$ (second row of Fig.~\ref{fig:gesamt}), $N=6$ (third row) and $N=12$ (last row) are shown.
These spectra are calculated for the same values of $V$ as the corresponding dimer spectra of the first row. 
Since in a cyclic aggregate with $N\ge 2$ each monomer possesses two nearest neighbors, for the same $V$ the effective coupling strength is now given by $C=2V$ leading to a shift of the mean of the absorption peak, that is twice as large as for the dimer~\cite{FuGo64_2280_,BrHe70_1663_,Ei07_321_}.   
Let us first consider the spectra for the case $N=3$ (Figs.~\ref{fig:gesamt}d-f).
Here a similar development of the width as for the dimer can be observed. 
The spectrum initially narrows very fast and reaches, for approximately $V=-2$, nearly its asymptotic line shape (as can be seen in Fig.~\ref{fig:gesamt}e).
However, the spectra are much narrower than the corresponding dimer spectra.

When considering a longer aggregate with $N=6$ (Figs.~\ref{fig:gesamt}g-i) we find that one has to go to much larger $|V|$ than in the case $N=3$ to reach the asymptotic line shape.
For $N=6$ it is  reached roughly for $V=-12$ (see Fig.~\ref{fig:gesamt}i).
For even larger aggregates larger values of $|V|$ are needed to achieve the asymptotic line shape as one can see in Fig.~\ref{fig:gesamt}j-l, for the case $N=12$.
Going from the top to the bottom row of Fig.~\ref{fig:gesamt} one observes that for the same $V$ but increasing $N$ the spectra also become significantly narrower.

\begin{figure}[tp]
\includegraphics[width=0.45\mylenunit]{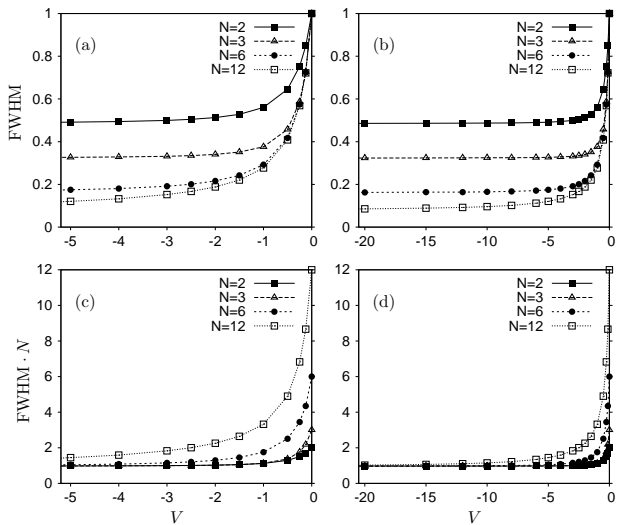}
\caption{\label{fig:FWHM} a) FWHM of $N$-mer absorption spectrum ($N$ indicated on the figures) over $V$; the corresponding monomer spectrum is that of Fig.~\ref{fig:monomer}f and the FWHM of the $N$-mer spectrum is given in units of the FWHM of the monomer spectrum.
c) $({\rm FWHM}\cdot N)$ over $V$.
b, d) Same as a, c but for a larger range of $V$.
The lines connecting the data points are only a guide for the eye.}
\end{figure}

To investigate quantitatively  the dependence of the narrowing of the spectrum on $N$ and $|V|$, we numerically evaluated its full width at half maximum (FWHM).
The results are shown in Figure~\ref{fig:FWHM}.
Here for different $N$ the FWHM is plotted against $V$.
To show the narrowing for small $|V|$ in closer detail, in the left column of Fig.~\ref{fig:FWHM} a smaller range of $V$ is taken.
While in the upper row the bare FWHM is plotted (in units of the FWHM of the
monomer spectrum), the lower row shows the FWHM multiplied by the number of monomers, i.e.\ ($N\cdot$FWHM). 
In Fig.~\ref{fig:FWHM}a,~b one observes that the larger $N$, the larger is the coupling strength $|V|$ needed for the FWHM curve to become horizontal, i.e.\ to reach the asymptotic FWHM.
This is consistent with the observation made by considering Figure~\ref{fig:gesamt}.
In Fig.~\ref{fig:FWHM}d one sees that the quantity ($N\cdot$FWHM) saturates for all $N$ at the FWHM of the monomer spectrum (which in Fig.~\ref{fig:FWHM} is taken as unity). 
That means that in the limit of strong interaction $V$, where the shape of the spectrum does not change anymore upon increasing $|V|$ further, the FWHM of the $N$-mer spectrum is narrowed by the factor $1/N$ w.r.t.\ the FWHM of the monomer spectrum.
This is a much stronger narrowing than the narrowing by the factor $1/\sqrt{N}$ found usually (the $1/\sqrt{N}$ narrowing has been found e.g.\ in the case of uncorrelated Gaussian disorder~\cite{Kn84_73_,FiKnWi91_7880_,Ma93_225_,WaEiBr08_044505_}, for both the FWHM and the standard deviation of the absorption peak).
However, it should be mentioned that in contrast to this narrowing of the FWHM
of the one dominant absorption peak by the factor $1/N$, as discussed above, the standard deviation of the spectrum only narrows by a factor $1/\sqrt{N}$.
A simple explanation for the narrowing of the FWHM will be given in the next section.

\subsection{Simple explanation of the anomalous strong narrowing}
\label{sec_explanation}

From analytic arguments presented in Ref.~\cite{WaEiBr08_044505_} one expects that for $|V|\gg \Delta$, where $\Delta$ denotes the width of the monomer absorption spectrum, the absorption spectrum of the interacting monomers (aggregate spectrum) has the same shape as the spectrum of the non-interacting monomers (monomer spectrum) obtained for a reduced coupling strength $\tilde{\kappa}_{\lambda}=\kappa_{\lambda}/\sqrt{N}$ to the environmental modes.
It follows from Eq.~(\ref{spec_dens}) that according to the reduced coupling
strength $\tilde{\kappa}_{\lambda}$ one has an effective spectral density
$\tilde{J}(\omega)=J(\omega)/N$.
This argumentation applies to arbitrary spectral densities, as long as
  the strong coupling condition $|V|\gg \Delta$ is fulfilled.

\begin{figure*}
\includegraphics[width=0.85\mylenunit]{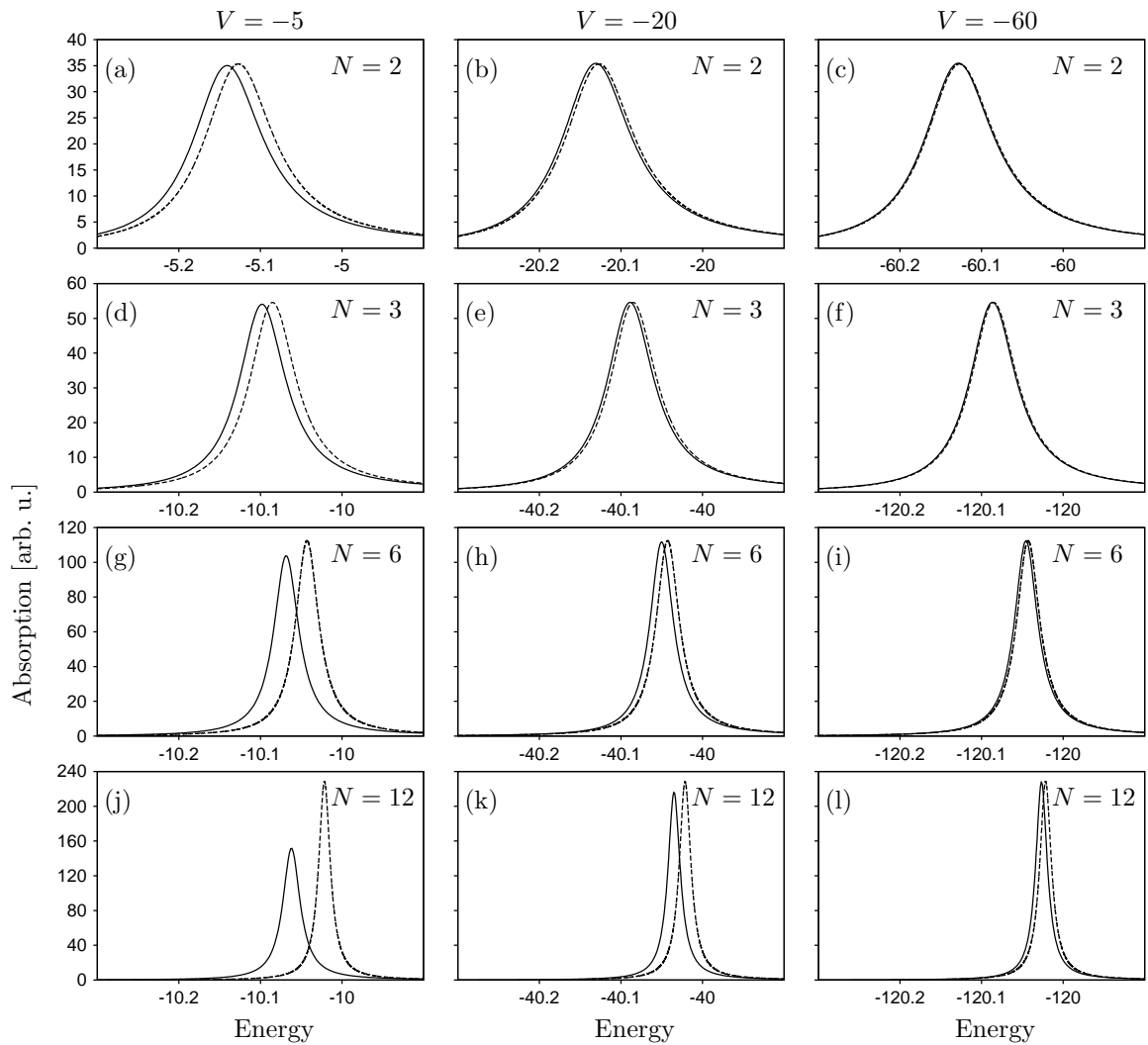}
\caption{\label{fig:comparison_w_analytic} 
$N$-mer spectra for large $|V|/\Delta$, where $\Delta$ is the width of the monomer spectrum. 
Comparison of numerically calculated (exact) $N$-mer spectra (solid lines) with the analytical monomer line shape obtained from the replacement $X\rightarrow X/N$ and shifted by $C$ (dashed).
Left column: $V=-5$, 
middle column: $V=-20$,
right column: $V=-60$.
Note the different energy and absorption scales.} 
\end{figure*}

For the Lorentzian spectral density Eq.~(\ref{lorentz_spec_dens}) that is considered here, this leads to a reduction of $X$ by a factor $1/N$, i.e.\ $\tilde{X}=X/N$.
Thus the aggregate spectrum in the limit of strong interaction $V$ can be obtained by calculating the monomer spectrum for $\tilde{X}=X/N$ and shifting it by $C$.
That this reasoning is indeed correct is demonstrated in Figure~\ref{fig:comparison_w_analytic}, where for the case of strong interaction $V$, i.e.\ large $|V|/\Delta$, numerically calculated (exact) $N$-mer spectra (solid lines) are compared with the analytic monomer line shape obtained for $\tilde{X}=X/N$ (dashed lines).
The analytic monomer spectra are shifted by $C=2V$ (or $C=V$ for the dimer respectively).
In Figure~\ref{fig:comparison_w_analytic} three different values of $V$ are
chosen, namely $V=-5$ (first column), $V=-20$ (second column) and $V=-60$
(third column).
For the dimer the agreement between numerical and analytic spectrum is already nearly perfect for $V=-20$, as can be seen in Fig.~\ref{fig:comparison_w_analytic}b.
However, for larger $N$ one has to go to larger $|V|$ to achieve perfect
agreement, shown for the cases $N=6$ and $N=12$ in the third and fourth row of Fig.~\ref{fig:comparison_w_analytic}.
This is due to the fact observed already in the previous section that for
larger $N$ a larger coupling strength $|V|$ is needed to reach the regime where the line shape has nearly converged, i.e.\ the strong coupling regime in which the arguments for the replacement $X\rightarrow X/N$ of Ref.~\cite{WaEiBr08_044505_} are valid.  
As described in Ref.~\cite{WaEiBr08_044505_} the standard deviation of the monomer spectrum with $\tilde{X}=X/N$ is reduced by a factor $1/\sqrt{N}$ compared to the original monomer spectrum with Huang-Rhys factor $X$.  

The narrowing of the FWHM of the aggregate spectrum by a factor $1/N$ observed
in Section~\ref{sec_lorentz_spec_dens} can now be explained as follows:
As is shown in the appendix, for the case $X\le 0.3$ considered above, only one Lorentzian-like peak, whose width is proportional to $X$, dominates the monomer spectrum and contributes to the FWHM.
Thus, when $X$ is decreased by $1/N$, also the FWHM of the monomer spectrum decreases by $1/N$.
Since the shape of the aggregate spectrum in the limit of strong interaction $V$ is that of the monomer spectrum with $X\rightarrow X/N$, the FWHM of the aggregate spectrum narrows by $1/N$.

\section{Conclusions}
\label{sec:conclusions}

We have investigated the exchange narrowing for an one-dimensional aggregate interacting with Markovian and non-Markovian environments.
In the case of a Markovian environment no narrowing occurred.
For a non-Markovian environment with a Lorentzian spectral density we found enhanced narrowing with a factor $1/N$ compared to the usual $1/\sqrt{N}$.
This usual $1/\sqrt{N}$ narrowing is found for uncorrelated Gaussian disorder (for the FWHM as well as for the standard deviation of the absorption peak), but also for the standard deviation of the aggregate spectrum for the present model Hamiltonian of Section~\ref{sec:model}, describing an exciton that couples to harmonic vibrations.
Thus, some comments about the apparent contradiction between the $1/\sqrt{N}$ narrowing, found e.g.\ in Ref.~\cite{WaEiBr08_044505_}, and the $1/N$ narrowing, found in the present work, are in order, since both rely on the same model Hamiltonian.
It is important to note that the extent of the narrowing depends crucially  on the definition of the ``width'' of the absorption spectrum.
For example, from a sum rule it follows exactly that the standard deviation of the absorption spectrum of the aggregate is independent of the interaction $V$, in particular it is identical to the standard deviation of the monomer spectrum. 
Thus, no narrowing occurs, if one takes the standard deviation, which is a global quantity of the whole spectrum, as a measure of the width.
It must be pointed out, that also weak absorption lines, well separated from the absorption peak one is interested in, contribute to the standard deviation.
In Ref.~\cite{WaEiBr08_044505_} such small contributions very far away from
the  absorption peak considered are neglected. 
One then finds that the coupling strength to each vibrational mode is
decreased by a factor $1/\sqrt{N}$, leading to an absorption line shape that
is a Poissonian, but with a standard deviation decreased by $1/\sqrt{N}$ w.r.t.\ the Poissonian of the monomer spectrum.
Thus, one obtains an overall $1/\sqrt{N}$ narrowing~-- again based on the standard deviation which gives large weight to small contributions far away from the mean.

In the present paper we have considered the full width at half maximum (FWHM) of an absorption peak, which is a local measure of the width. 
It is applicable whenever there is a well defined peak, as in the present study.
By choosing this measure we neglect the wing at high energies, which can be seen in all spectra and which contributes the additional width when the standard deviation is considered.   

Furthermore, it should be mentioned that for larger Huang-Rhys factors, when
more than one peak of the vibrational progression in the monomer spectrum
contribute to the FWHM, the narrowing of the FWHM will not have the simple
$1/N$ dependence anymore (see appendix).

While in this work we have focused on a Lorentzian spectral density,
  the main results, namely an enhanced narrowing compared to the
  $1/\sqrt{N}$ case, are expected to remain valid also for other forms of the
  spectral density (as discussed in Section~\ref{sec_explanation}).
We have performed calculations also for an ohmic spectral density with exponential cutoff and found similar results as discussed here for the Lorentzian spectral density. 

The findings of this work demonstrate, that one should be careful when
deducing quantities such as the number of coherently interacting monomers from
the narrowing of the absorption spectrum, since the exact value of the
narrowing strongly depends not only on the model assumed (e.g.\ static
  disorder, single vibrations, a continuous spectral density, etc.) but also on
the definition of the width.

\appendix

\section{Analytical formula for the monomer absorption spectrum}
\label{sec:mono_analyt}

The monomer absorption spectrum is given by (cf.\ Eq.~(\ref{eq:abs_spek_mit_tkorr_fkt}))
\begin{equation}
\label{mon_abs_spec}
A(\LichtFr)=\Real \int_0^{\infty} dt \ e^{i \LichtFr t}\ c(t),
\end{equation}
where the correlation function $c(t)= \bra{\Psi_0}{\rm e}^{-iHt}\ket{\Psi_0}$ can be determined analytically for various concrete bath correlation functions $\alpha(\tau)=\int d\omega\, e^{-i\omega\tau}J(\omega)$ (here for zero temperature; $J(\omega)$ is the spectral density of the environmental modes).
In the case of the Lorentzian spectral density Eq.~(\ref{lorentz_spec_dens}), that we consider in this work, we have
\begin{equation}
\label{BathCorrFuncForLorentzSpecDens}
 \alpha(\tau)=\Gamma\ e^{-i \Omega\tau - \gamma\tau}
\end{equation}
with $\Gamma=\Omega^2 X$ (note that since in this work we always take $\Omega=1$, the parameter $\Gamma$ is equal to the Huang-Rhys-factor $X$).
For this $\alpha(\tau)$ the correlation function $c(t)$ needed in
Eq.~(\ref{mon_abs_spec}) can be evaluated analytically to give~\cite{footnotes}
\begin{equation}
\label{dip_corr_fct_for_lorentz_spd}
  c(t)=\exp\left\{-i\varepsilon t-\frac{\Gamma}{w^2}\left(wt+e^{-wt}-1\right)\right\},
\end{equation}
where $\varepsilon$ is the transition energy of the monomer and with $w=i\Omega+\gamma$.
Inserting Eq.~(\ref{dip_corr_fct_for_lorentz_spd}) into Eq.~(\ref{mon_abs_spec}) leads to an analytic expression for the absorption spectrum
\begin{equation}
\label{analyt_mon_spec}
A(\LichtFr)= e^{(\tilde{\gamma}^2-\tilde{\Omega}^2)/\Gamma}
\sum_{k=0}^{\infty} \frac{A_k(\LichtFr)}{k!}
\Big(-\frac{\tilde{\gamma}}{\gamma}\,{\rm sign}(\gamma^2-\Omega^2)\Big)^k,
\end{equation}
with the definitions 
\begin{eqnarray}
\tilde{\gamma}&=& \Gamma\gamma/(\gamma^2+\Omega^2)\\
\tilde{\Omega}&=& \Gamma\Omega/(\gamma^2+\Omega^2)
\end{eqnarray}
and with
\begin{equation}
A_k(\LichtFr)=\cos(\theta-kq)\ {\rm Im}\,L_k(\LichtFr)+\sin(\theta-kq)\ {\rm Re}\, L_k(\LichtFr),
\end{equation}
where $\theta=2 \tilde{\gamma} \tilde{\Omega}/\Gamma$ and $q=\arctan(-2 \gamma \Omega/(\gamma^2-\Omega^2))$.
Here
\begin{equation}
L_k(\LichtFr)=\frac{1}{(\LichtFr - k\Omega + \tilde{\Omega}-\varepsilon) - i
(\tilde{\gamma}+k\gamma)}
\end{equation}
is a complex Lorentzian line shape function.
The spectrum Eq.~(\ref{analyt_mon_spec}) consists of a sum of terms where the line shape of each term is determined by $L_k(\LichtFr)$. 
The imaginary part of $L_k(\LichtFr)$ is a Lorentzian centered at $(k\Omega-\tilde{\Omega}+\varepsilon)$ and with a width $(\tilde{\gamma}+k\gamma)$. 
Thus the Lorentzians for the different $k$ are spaced with a distance $\Omega$ and their width increases with increasing $k$.
The real part of $L_k(\LichtFr)$ is given by the same Lorentzian but multiplied with the linear function $(\nu-k\Omega+\tilde{\Omega}-\varepsilon)/(\tilde{\gamma}+k\gamma)$.
  
Since the first peak ($k=0$) is of particular interest in this paper (because it dominates the spectrum for small $\Gamma$) we discuss it in more detail. 
Denoting the corresponding term in the sum in Eq.~(\ref{analyt_mon_spec}) by $A_0(\LichtFr)$ we have
\begin{equation}
A_0(\LichtFr)=
\frac{\tilde{\gamma}}{(\LichtFr +\tilde{\Omega}-\varepsilon)^2+\tilde{\gamma}^2}
\left(\cos(\theta)+\sin(\theta)\frac{\LichtFr +\tilde{\Omega}-\varepsilon}{\tilde{\gamma}}\right).
\end{equation}
For the values taken in the investigations of Section~\ref{sec:results}, i.e.\ $\Gamma\le 0.3$ and $\gamma=0.4$, the line shape of this peak is to a good approximation a Lorentzian with a width (HWHM) given by $\tilde{\gamma}$.
Since $\tilde{\gamma}\propto \Gamma$, replacing $\Gamma$ by $\Gamma/N$ to obtain the line shape of the $N$-mer in the strong coupling limit (as has been done in Section~\ref{sec_explanation}) one finds a narrowing of the $N$-mer peak by a factor $1/N$ w.r.t.\ the monomer. 
However, for a stronger coupling $\Gamma$ to the environment, when beside the $k=0$ peak also other peaks contribute to the FWHM of the spectrum, the narrowing of the FWHM does not have the simple $1/N$ dependence anymore.

\begin{acknowledgments}
We thank Walter T.\ Strunz for many helpful discussions and John S.\ Briggs for
commenting on the manuscript.
\end{acknowledgments}


\end{document}